# Blueshift of plasmon resonance with decreasing cluster size in Au nanoclusters embedded in silica matrix

S. Dhara,[*] B. Sundaravel, T. R. Ravindran, K. G. M. Nair, B. K. Panigrahi, and

P. Magudapathy

Materials Science Division, Indira Gandhi Centre for Atomic Research, Kalpakkam –603 102,

India

## Abstract

Gold nanoclusters are grown by 1.8 MeV $Au^{++}$ implantation in silica matrix and subsequent air annealing in the temperature range of 873K-1273K. Post-annealed samples show plasmon resonance in the optical region (~ 2.38-2.51 eV) for average cluster radii ~1.04-1.74 nm. A blueshift of the plasmon peak is observed with decreasing cluster size in the annealed samples. Similar trend of blueshift with decreasing cluster size in case of Au nanoclusters embedded in the porous alumina matrix [B. Palpant, *et al.*, *Phys. Rev.* B **57**, 1963 (1998)] has convinced us to assume a possible role of 'rind' like porosity at the Au nanocluster-matrix interface with available open volume defects in the amorphous silica matrix.

________

Electronic mail : dhara@igcar.ernet.in

Noble metals embedded in dielectric matrices earned tremendous importance in last few years with applications in non-linear optics,[1,2] and understanding of various fundamental issues related to the surface plasmon resonance (SPR, collective oscillation of the conduction electrons) in the UV-Visible range.[3-5] The observation of SPR peaks in the visible region gives an extra impetus to various optical-device applications. Enormous interest is there to understand the shift of the SPR peaks with cluster size as many a fundamental issue is involved in it, e.g., a redshift is observed in small clusters owing to reduction in the electron density ('spillout' effect)[4,6] and a blueshift for larger clusters is recognized mainly due to the localization of conduction electrons.[3,4] Thus, measuring the absorption spectra of an ensemble of embedded nanoclusters is an indirect method of determining their size. A large number of experimental and theoretical studies reported the blue shift of the SPR peak position with decreasing noble metal cluster size as an effect of embedding matrix,[3-5] and of the surrounding porosity.[3,4,7] An experimental observation of blueshift of SPR peak position with decreasing size is reported for Au (2-4 nm) clusters embedded in porous alumina matrix grown by co-deposition technique using pulsed laser ablation.[7] Effect of porosity surrounding Au clusters is taken into consideration in a time dependent local density approximation (TDLDA)[3,4] calculation to support the experimental observation.

Large amount of studies are performed for the growth and properties of Au nanoclusters in dielectric matrices, as these clusters are preferred over other noble metals for stable and quality device structure. Among various matrices the most common is silica, which finds applications in the optical devices. Au clusters in silica are formed by various techniques which includes chemical routes,[8] co-sputtering,[9] and ion implantation techniques.[10-12] However, ion beam technique for the formation of nanoclusters has certain advantages over other techniques:

excellent control of size, selectivity of depth, and area in which the nanoclusters form.[13] In the present study, we report the growth of Au nanoclusters in fused quartz (silica) matrix using direct Au ion-implantation at various fluence ranges and subsequent annealing at high temperatures. A blueshift of the SPR peak position with decreasing cluster size is reported in the cluster (radii ~1.04-1.74 nm) embedded in the amorphous silica matrix.

Au clusters were grown by direct Au-ion implantation on silica substrates. Implantation was performed using 1.8 MeV $Au^{++}$ at $1 \times 10^{-5}$ Pa with a beam current of ~12 mA $m^2$ to avoid substantial beam heating in the ion fluence range of $2 \times 10^{20}$-$1 \times 10^{21}$ $m^{-2}$. A 1.7 MV Tandetron accelerator (High Voltage Engineering Europa, The Netherlands) was employed for the implantation study. High temperature air annealing in the range of 873K-1273K for 1 hour was performed for studying the growth of the clusters. Optical absorption spectra were recorded at room temperature in the range of 200-1100 nm using a Hewlett-Packard diode-array UV-visible (UV-Vis) spectrophotometer (Model 8453) with necessary signal correction for the substrate. Cluster sizes were determined using low-frequency Raman scattering studies at room temperature in the back scattering geometry, using vertically polarized 488 nm line of an argon ion laser (Coherent, USA) with 200 mW power. Unpolarized scattered light from the sample was dispersed using a double monochromator (Spex, model 14018) with instrument resolution of 1.4 $cm^{-1}$ and detected using a cooled photomultiplier tube (FW ITT 130) operated in the photon counting mode.

Plasmon peaks around ~ 2.38 - 2.51 eV are observed in the UV-Vis absorption study of the annealed samples grown at different fluences (Fig. 1). We could not observe any SPR peak in the as-grown samples. However, SPR peaks were observed for the annealed samples where cluster may have grown into a sufficiently big size (to be discussed later) for observable resonance to occur. The surface plasmon resonance (SPR) frequencies in the large sized clusters is estimated at $\omega_{SPR}(\omega) = \omega_P/[2\varepsilon_m(\omega) + Re(\varepsilon_d(\omega))]^{1/2}$ where $\omega_P$ is the bulk plasmon frequency. $\varepsilon_d(\omega) = 1 + \chi^d$ is the core-electron contribution to the complex dielectric function of the noble metal $\varepsilon(\omega) = 1 + \chi^s(\omega) + \chi^d(\omega)$ [$\chi^d$ the interband part of dielectric susceptibility (*d* electrons); $\chi^s$ the Drude-Sommerfeld part of dielectric susceptibility (*s* electrons)]. $\varepsilon_m$ (1 in vacuum and $\approx 2.16$ for silica in the relevant energy range)[14] is the dielectric function of matrix. The Mie frequency ($\omega_{SPR}$) in the large-size limit for embedded Au nanoclusters in silica is calculated by solving $\varepsilon(\omega) + 2\varepsilon_m(\omega) = 0$ and found to be ~ 2.37 eV. This value is close to our observed values of SPR peaks (Fig. 1) indicating formation of Au nanoclusters. Irrespective of growth fluences we observed a blueshift of the SPR peak position with decreasing annealing temperatures (Fig. 1).

Low-frequency Raman study was performed for the determination of size and shape of clusters in the post-annealed samples. Confined surface acoustic phonons in metallic or semiconductor nanoclusters give rise to low-frequency modes in the vibrational spectra of the materials. The Raman active spheroidal motions are associated with dilation and strongly depend on the cluster material through $v_t$, the transverse and $v_l$, the longitudinal sound velocities. These modes are characterized by two indices *l* and *n*, where *l* is the angular momentum quantum number and *n* is the branch number. $n = 0$ represents surface modes. The surface quadrupolar mode $l = 2$, eigenfrequency $\eta^s_2$ appears in Raman scattering geometries whereas, the surface spherical mode $l = 0$, eigenfrequency $\xi^s_0$ appears only in the polarized geometry. Eigenfrequencies for the spheroidal modes on the surface ($n = 0$; $l = 0, 2$) of Au nanocluster in silica matrix is calculated to be $\eta^s_2 = 0.67$ and $\xi^s_0 = 0.34$ by considering the matrix effect in the limit of elastic body approximation of small cluster.[15] The surface quadrupolar mode frequencies corresponding to $l = 0$, and 2 are given by,

$$\omega^s_0 = \xi^s_0 \, v_l / Rc \, ; \, \omega^s_2 = \eta^s_2 \, v_t / Rc \ldots\ldots\ldots\ldots(1)$$

where $c$ is the velocity of light in vacuum, and $v_l$ = 3240 m/sec, $v_t$ = 1200 m/sec in Au. Figures 2(a)-(c) show the Raman spectra of post-annealed samples prepared at various fluences. Average cluster sizes ($R \approx$ 1.04 -1.74 nm $\approx$ 19.7-32.9 a.u.), calculated using Eqn. (1), corresponding to $l$ = 0, 2 are found to be nearly the same as expected and the mean values are inscribed in the Figs. 2(a)-(c) for the corresponding fluences at different annealing conditions. The increase in the cluster size with increasing fluence and annealing temperature can be understood as follows. In the first stage, fluctuations in concentration nucleate gold atoms and these nuclei can grow directly from the supersaturation. Implanted Au is supersaturated in the silica matrix due to its low solubility. Thus, with increasing fluence, agglomeration to bigger clusters is likely.[16] In the second stage, clusters are likely to grow with increasing annealing temperature and this may be due to the fact that nuclei, which have reached a critical size, grow at the expense of smaller nuclei and known as the coarsening or ripening stage.[16] Thus, the cluster size should be larger with higher fluence and for a particular fluence the cluster size is expected to increase with increasing annealing temperature [Figs. 2(a)-(c)].

For these nanoclusters in the size range of ~1.04-1.74 nm [inscribed in Figs. 2(a)-(c)], a blueshift is observed with decreasing cluster size (Fig. 3). We have compared our results with Au nanoclusters embedded in porous alumina matrix.[7] The blueshift of SPR peak energy with decreasing size follows the similar trend (Fig. 3) of Au nanoclusters embedded in alumina matrix. Normally a redshift is expected for very small clusters with decreasing cluster size due to electron 'spillout' effect.[4,6] However, as predicted by a time dependent local density approximation (TDLDA) calculation,[3,4] with increasing cluster size the 'spillout' effect is overcome by the steep frequency dependence of $\varepsilon_d(\omega)$ near $\omega_{SPR}$ arising from core electron polarization effect, especially of electrons belonging to the fully occupied *d* band e.g., noble metals. As discussed earlier, cluster-matrix interface effect is also reviewed under TDLDA formalism,[3,4] and a pronounced blueshift with decreasing cluster size is demonstrated to explain the observation for Au nanoclusters embedded in alumina matrix with 'rind'-like porosity (separation ~ 2 a.u.) at the interface.[3,4] In the present study, open volume defects in the amorphous silica matrix is sufficient to indicate the possible existence of a 'rind'-like porosity (separation ~ 1-2 a.u.)[17] at the cluster-matrix interface and that can be accounted for the observed blueshift.

In conclusion, embedded Au nanoclusters of radii ~ 1.04-1.74 nm were grown by direct ion implantation in silica matrix and subsequent air annealing treatment. Plasmon resonance spectra were recorded in the optical region of ~ 2.38 – 2.51 eV. A blueshift of the plasmon peak with decreasing cluster size was observed. The blueshift is correlated to the porosity effect for the embedded Au clusters in the amorphous silica matrix.

The authors like to thank V. K. Indirapriyadarshini, and P. Ramamurthy of National Centre for Ultrafast Processes, University of Madras, India for their co-operation in UV-Vis studies. We thank C. David and P. Gangopadhyay of MSD, IGCAR for their contributions in the implantation work. We also thank B. Viswanathan of MSD, IGCAR for his encouragement in pursuing this work.

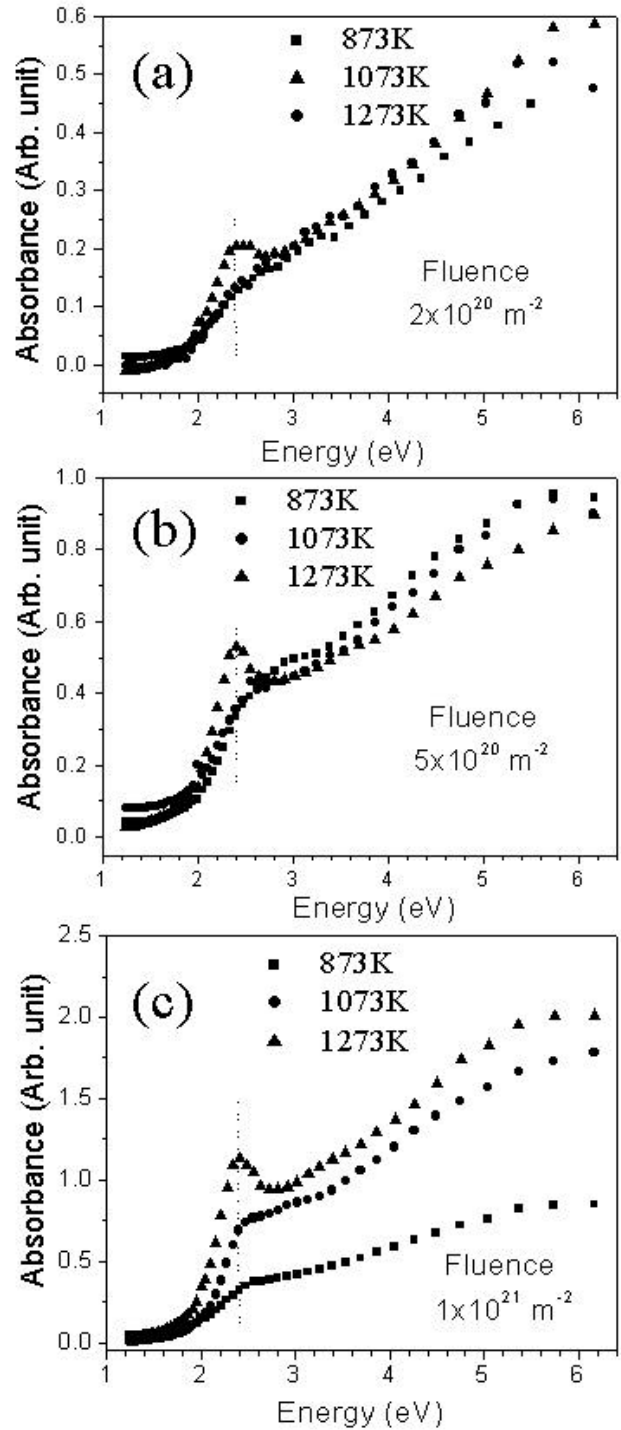

Fig. 1. SPR peaks for post-annealed Au nanoclusters in silica matrix grown at fluences (a) $2\times10^{20}$ $m^{-2}$ (b) $5\times10^{20}$ $m^{-2}$ and (b) $1\times10^{21}$ $m^{-2}$. Dashed vertical line is a guide to eye for the observed blueshift of the SPR peak with decreasing annealing temperature.

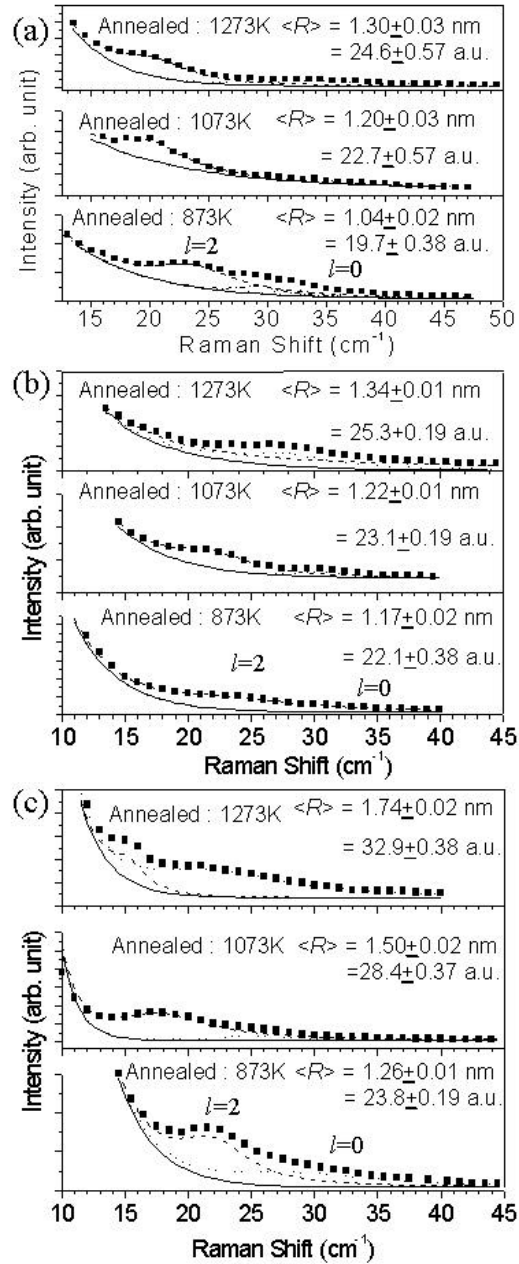

Fig. 2. Low-frequency Raman spectra for post-annealed Au nanoclusters in silica matrix grown at fluences of (a) $2\times10^{20}$ m$^{-2}$, (b) $5\times10^{20}$ m$^{-2}$, and (b) $1\times10^{21}$ m$^{-2}$. Average cluster sizes (radii, $R$) determined from the eigenfrequencies corresponding to l = 0 and 2 (refer text) are inscribed for the corresponding fluences [Background (continuous lines); l = 0 (dotted lines) and l = 2 (dashed and dashed-dot lines)]. The radii are also presented in atomic unit (a.u.~ 0.0529 nm).

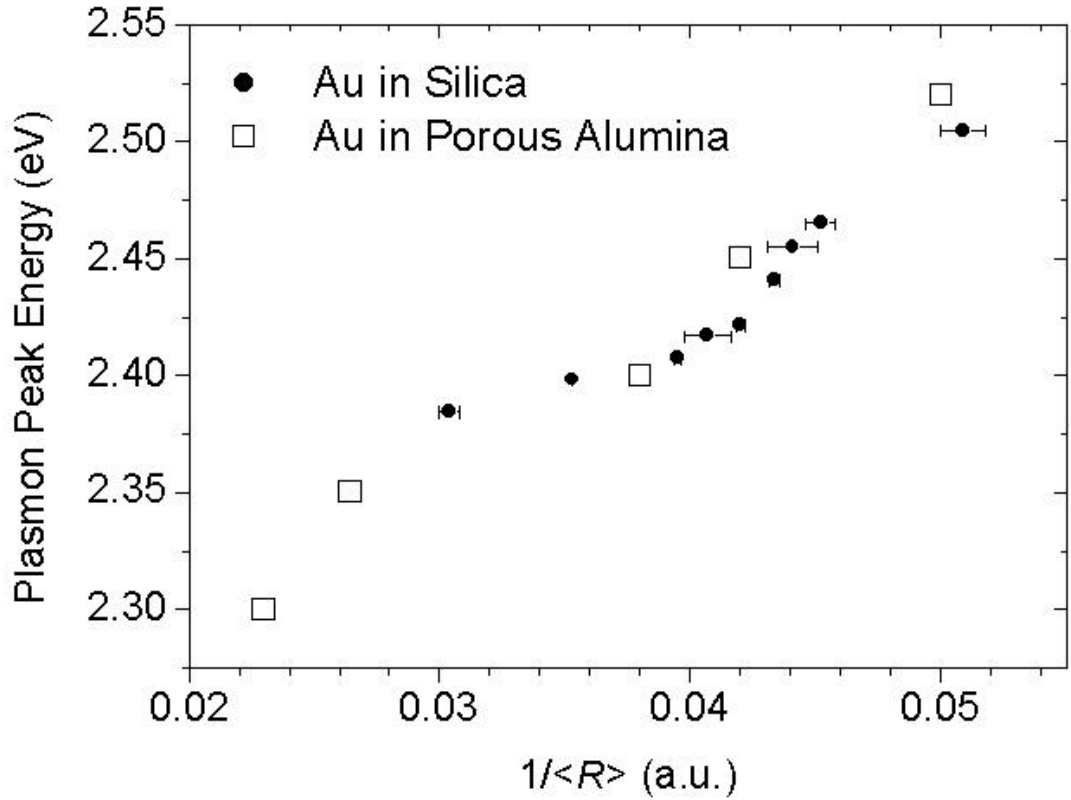

Fig. 3. Size dependence of the plasmon resonance peak for the post-annealed Au nanoclusters in silica matrix grown in the fluence range of $2 \times 10^{20}$-$1 \times 10^{21}$ m$^{-2}$ (filled symbols) and for the Au nanoclusters in porous alumina matrix (unfilled symbols) [Refs. 3,4,7]. Both the results show a similar trend of blueshift with decreasing cluster size.